\documentclass[12pt,journal,onecolumn]{IEEEtran}
\usepackage{setspace}
\usepackage{amsfonts,amssymb}
\usepackage{amsthm}
\usepackage{graphicx,graphics}
\usepackage{cite,url}
\usepackage{flushend}
\usepackage{smartref}
\usepackage{amsmath}
\usepackage{xcolor}

\usepackage{tikz}
\allowdisplaybreaks
\newcommand{\rmT}{\mathrm{T}}
\newcommand{\rmR}{\mathrm{R}}
\newcommand{\rmB}{\mathrm{B}}
\newcommand{\rmH}{\mathrm{H}}
\newcommand{\rmI}{\mathrm{I}}
\newcommand{\rme}{\mathrm{e}}
\newcommand{\rms}{\mathrm{s}}
\newcommand{\rmt}{\mathrm{t}}
\newcommand{\rmd}{\mathrm{d}}

\begin{document}
	\title{Localization Error Bounds For 5G mmWave Systems Under I/Q Imbalance: An Extended Version}
	\author{
		Fariba Ghaseminajm,
		Zohair Abu-Shaban~\IEEEmembership{Senior Member, IEEE}, \\ 	
		Salama S. Ikki~\IEEEmembership{Senior Member, IEEE},
		Henk Wymeersch~\IEEEmembership{Senior Member, IEEE}, and
		Craig R. Benson~\IEEEmembership{Member, IEEE}.
	}
	
	\maketitle
	
	\thispagestyle{empty}
	\let\thefootnote\relax\footnotetext{
		Fariba Ghaseminajm and Salama S. Ikki are with the Department of Electrical Engineering, Faculty of Engineering, Lakehead University, Thunder Bay, Ontario, Canada. Emails: \{fghasemi, sikki\}@lakeheadu.ca. Zohair Abu-Shaban and Craig R. Benson are with the School of Engineering and Information Technology, University of New South Wales (UNSW), Canberra, Australia. Emails: \{zohair.abushaban, c.benson\}@unsw.edu.au. Henk Wymeersch is with the Department of Signals and Systems, Chalmers University of Technology, Sweden. Email: henkw@chalmers.se.
	}
	\begin{abstract}
		Location awareness is expected to play a significant role in 5G millimeter-wave (mmWave) communication systems. One of the basic elements of these systems is quadrature amplitude modulation (QAM), which has in-phase and quadrature (I/Q) modulators. It is not uncommon for transceiver hardware to exhibit an imbalance in the I/Q components, causing degradation in data rate and signal quality. Under an amplitude and phase imbalance model at both the transmitter and receiver, 2D positioning performance in 5G mmWave systems is considered. Towards that, we derive the position and orientation error bounds and study the effects of the I/Q imbalance parameters on the derived bounds. The numerical results reveal that I/Q imbalance impacts the performance similarly, whether it occurs at the transmitter or the receiver, and can cause a degradation up to 12\% in position and orientation estimation accuracy.
				\vspace{-5mm}
	\end{abstract}
	
	\section{Introduction}
	Millimeter-wave (mmWave) systems is a major topic contributing to enhancing the fifth generation (5G) mobile communication systems. They offer high bandwidth, leading to higher data rates, and use carrier frequencies from 30 GHz to 300 GHz \cite{overview}. In parallel, location-aided systems in 5G are numerous and serve in a wide range of applications such as vehicular communications and beamforming. 
	
	Due to the employment of antenna arrays at both the base station (BS) and user equipment (UE), single-anchor localization through the estimation of the directions of arrival and departure (DOA, DOD) and the time of arrival (TOA) is possible. Single-anchor localization bounds for 5G mmWave systems have been widely considered  in the literature. For example, in \cite{Zohair2017}, the 3D position error (PEB) and the orientation error bounds (OEB) have been studied for uplink and downlink localization, while \cite{Arshpaper} proposed position and orientation estimators for 2D positioning. In \cite{Lone2019}, the authors investigated the probability of 5G localization with non-line-of-sight paths, while \cite{multipatheffect} investigated localization bounds in multipath MIMO systems. 
	
	Quadrature amplitude modulation (QAM) is widely used in modern communication systems, particularly mmWave systems. In this modulation, in-phase (I) and quadrature (Q) components should be perfectly matched. However, due to limited accuracy in practical systems, a perfect match is rarely possible, leading to performance degradation, including positioning. Although the effect of IQ imbalance (IQI) on positioning was studies previously in several papers (See for example \cite{TRIPS}),  to the best of our knowledge, it has not been investigated for 5G, despite its severity in mmWave systems \cite{Hardware}. IQI gain and phase parameters are usually compensated during the channel estimation phase \cite{TRIPS}, \cite{IQI_channel}. In mmWave systems, this is the phase during which DOD, DOA, and TOA are estimated and ultimately a position fix is obtained. This implies that investigating IQI jointly with localization is crucial in the context of 5G mmWave systems.
	
	In this paper, we consider 2D mmWave uplink localization under IQI, focusing on the RF phase-shifting model \cite{phased_array}. To this end, we consider gain and phase imbalance at both the transmitter and receiver and derive the PEB and OEB. Subsequently, we investigate the resulting PEB and OEB degradation and obtain insights through numerical simulation.
	
	\section{Problem Formulation}
	\begin{figure}[!t]
		\centering
		\includegraphics[scale=1.0]{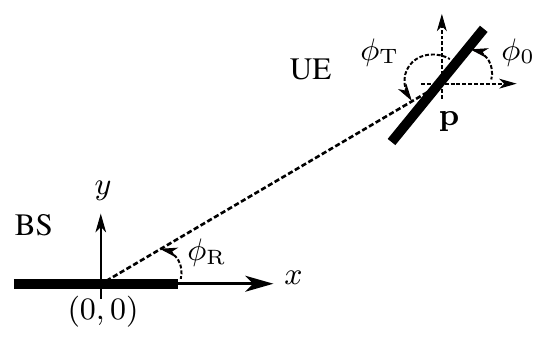}
		\caption{Considered geometry including UE and BS equipped with of ULAs with $N_\rmT$ and $N_\rmR$ antennas respectively.}
		\label{2DUEBS}
	\end{figure}
	
	Consider an uplink transmission scenario in which a BS is equipped with $N_\rmR$-antenna uniform linear array (ULA) lying on the $x$-axis and centered at the origin. The BS receives a signal from a UE with an $N_\rmT$-antenna ULA and an unknown orientation angle $\phi_0$ measured from the positive $x$-axis, as shown in Fig.~\ref{2DUEBS}. We assume that the UE location, $\mathbf{p}=[p_x, p_y]^\rmT $, to be unknown. We assume one path between BS and UE as line of sight (LOS) channel. Note that in the case of multipath, the LOS provides the highest useful information in terms of positioning \cite{Lone2019}, and due to path orthogonality \cite{Zohair2017}, it is easy to isolate it based on the received power profile.
	\subsection{Signal Model}
	\begin{figure*}[!t]
		\centering
		\vspace{-5mm}
		\includegraphics[scale=1]{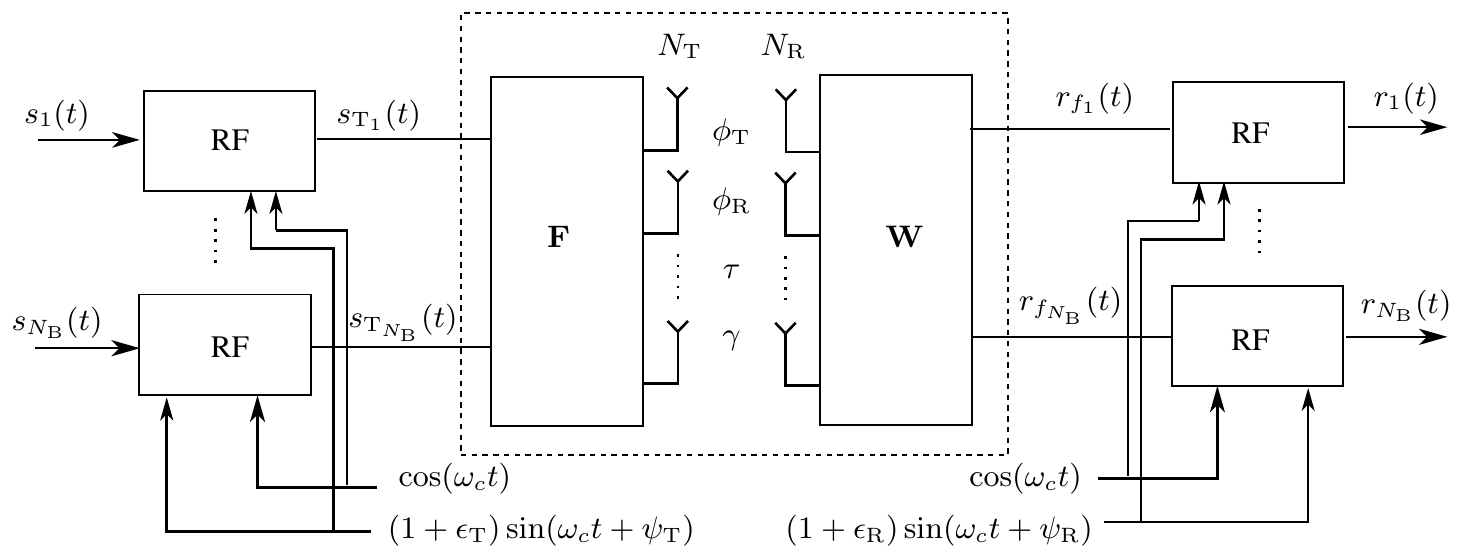}
		\caption{5G mmWave Transceiver structure under RF phase-shifting configuration ($\omega_c=2\pi f_c$ where $f_c$ is the carrier frequency)} 
		\label{TXRX}
	\end{figure*}
	The considered transceiver structure under I/Q mismatch is shown in Fig.~\ref{TXRX}. Based on \cite{IQ} and \cite{IQ.dr.ikki}, the baseband signal, $\mathbf{s}_\rmT(t)\triangleq [s_{\rmT_1}(t),\cdots, s_{\rmT_{N_\rmB}}(t)]^\rmT$, at the output of the RF chain can be written as
	\begin{align}\label{stt}
	\mathbf{s}_\rmT(t)=\alpha_\rmT \mathbf{s}(t)+\beta_\rmT \mathbf{s}^* (t),
	\end{align}
	where  $\mathbf{s}(t)\triangleq [s_{1}(t),\cdots, s_{N_\rmB}(t)]^\rmT$ is the baseband signal containing signals drawn from a zero-mean constellation and passed through a pulse shaping filter with PSD $P(f)$, $N_\rmB$ is the number of transmitted beams and 
	\begin{subequations}
		\begin{align}\label{alpha_beta_t}
		\alpha_\rmT\triangleq \frac 12 \left(1+m_\rmT e^{j\psi_\rmT}\right),\\
		\beta_\rmT\triangleq \frac 12 \left(1-m_\rmT e^{j\psi_\rmT}\right),
		\end{align}
	\end{subequations}
	such that $m_\rmT\triangleq 1+\epsilon_\rmT$ and $\epsilon_\rmT$ and $\psi_\rmT$ represent the amplitude and phase imbalance parameters of the transmitter outlined in Fig.~ \ref{TXRX}. Note that $E_{\mathrm{s}}$, the transmitted energy per symbol of $\mathbf{s}_\rmT(t)$, is related to $E_\rmt $, the energy per symbol of $\mathbf{s}(t)$, by
	\begin{align} \label{energy_growth}
	E_{\mathrm{s}}=\frac{2E_\rmt }{1+m_\rmT^2},
	\end{align}
	indicating that transmitter IQI leads to energy loss.

	Denoting the DOD, DOA  and propagation delay by $\phi_\rmT, \phi_\rmR$ and $\tau$, respectively, a widely used model (e.g., \cite{overview,Zohair2017}) to describe the input/output relationship of the dashed box in Fig.~\ref{TXRX} is
	\begin{align}
	\mathbf{r}_\mathrm{f}(t)\triangleq &\sqrt{E_{\mathrm{s}}N_\rmR N_\rmT}\gamma \mathbf{W}^\rmH \mathbf{a_\rmR}(\phi_\rmR)\mathbf{a_\rmT}^\rmH(\phi_\rmT)\mathbf{F}\mathbf{s}_\rmT(t-\tau)\notag\\
	&+\mathbf{W}^\rmH \mathbf{n}(t),\in C^{N_\rmB},\label{rf}
	\end{align}
	where $\gamma\triangleq\gamma_\rmR+j\gamma_\rmI$ is the complex path gain, and
	\begin{align}\label{at}
	\mathbf{a}_\rmT(\phi_\rmT)&=\frac{1}{\sqrt{N_\rmT}}e^{-j\frac{2\pi d}{\lambda}\cos\phi_\rmT\mathbf{x}_\rmT},
	\end{align}
	is the transmit array response vector, $d$ is the inter-element spacing, and  $\mathbf{x}_\rmT\triangleq \left[-\frac{N_\rmT-1}{2},-\frac{N_\rmT-1}{2}+1,...,\frac{N_\rmT-1}{2}\right]$ is the antenna location vector.  $\mathbf{a}_\rmR(\phi_\rmR)$ can be similarly defined. $\mathbf{F}=[\mathbf{f}_1,\cdots,\mathbf{f}_{N_\rmB}]\in\mathbb{C}^{N_\rmT\times{N_\rmB}}$ and  $\mathbf{W}=[\mathbf{w}_1, \cdots,\mathbf{w}_{N_\rmB}]\in\mathbb{C}^{N_\rmR\times{N_\rmB}}$ are the $N_\rmB$-beam analog transmit and receive beamforming matrices, respectively. Furthermore $\mathbf{n}(t)\triangleq [n_1(t), n_2(t),...,n_{N_\rmR}(t)]^\rmT  \in\mathbb{C}^{N_\rmR}$ denotes zero-mean additive white Gaussian noise with spectral density $N_0$, with independent real and imaginary parts.
	
	Similar to the transmitter side, taking $m_\rmR\triangleq 1+\epsilon_\rmR$, then based on \cite{IQ} and \cite{IQ.dr.ikki}, the received baseband signal, $\mathbf{r}(t)\triangleq [r_{1}(t),\cdots, r_{N_\rmB}(t)]^\rmT$, is
	\begin{align}
	\mathbf{r}(t)=&\alpha_\rmR\mathbf{r}_\mathrm{f}(t)+\beta_\rmR\mathbf{r}_\mathrm{f}^*(t),\label{r(t)}
	\end{align}
	where the receiver IQI parameters are defined as
	\begin{subequations}\label{rx_iq}
		\begin{align}\label{alpha_beta_r}
		\alpha_\rmR\triangleq&\frac 12 \left(1+m_\rmR e^{-j\psi_\rmR}\right),\\
		\beta_\rmR\triangleq&\frac 12 \left(1-m_\rmR e^{j\psi_\rmR}\right).
		\end{align}
	\end{subequations}
	\subsection{2D localization problem}
	Our goal is to obtain the UE PEB and OEB using the received signal, $\mathbf{r}(t)$. We achieve this in two steps: first, we derive Fisher information of channel parameters $\boldsymbol{\varphi}_\mathrm{C}\triangleq \{\phi_\rmR,\phi_\rmT,\tau,\gamma_\rmR,\gamma_\rmI,\epsilon_\rmR, \epsilon_\rmT,\psi_\rmR,\psi_\rmT\}$. Then, we transfer this Fisher information into the position domain using a transformation of parameters. 	
	
	\section{FIM of Channel Parameters}
	We now derive the Fisher Information Matrix (FIM) of the vector of observed parameters. Namely, define
	\begin{align} 
	\boldsymbol{\varphi}_\mathrm{C}\triangleq [\phi_\rmR,\phi_\rmT,\tau,\gamma_\rmR,\gamma_\rmI,\epsilon_\rmR, \epsilon_\rmT,\psi_\rmR,\psi_\rmT]^\rmT,
	\end{align}
	then, the corresponding FIM is denoted by
	\begin{equation}\label{jphiph}
	\mathbf{J}_\mathrm{C}=\begin{bmatrix} 
	{J}_{\phi_\rmR\phi_\rmR} &{J}_{\phi_\rmR\phi_\rmT}&\cdots&{J}_{\phi_\rmR\psi_\rmT}\\
	{J}_{\phi_\rmT\phi_\rmR} &\ddots&\cdots&{J}_{\phi_\rmT\psi_\rmT}\\
	\vdots & \vdots & \ddots & \vdots\\
	{J}_{\psi_\rmT\phi_\rmR} &\cdots&\cdots&{J}_{\psi_\rmT\psi_\rmT}\\
	\end{bmatrix} 
	\in\mathbb{R}^{9\times{9}}.
	\end{equation}
	The derivation of the elements in \eqref{jphiph} depends on whether the noise covariance matrix is a function of the parameter in question \cite{Statistical}. Therefore, we digress to compute the noise covariance matrix as follows.
	
	Taking $\mathbf{r}_\mathrm{o}(t-\tau)\triangleq\gamma \mathbf{W}^\rmH \mathbf{a_\rmR}(\phi_\rmR)\mathbf{a_\rmT}^\rmH(\phi_\rmT)\mathbf{F}\mathbf{s}_\rmT(t-\tau)$, based on \eqref{rf} and \eqref{r(t)}, we can write  
	\begin{align}
	\mathbf{r}(t)=& \underbrace{\sqrt{E_{\mathrm{s}}N_\rmR N_\rmT}\left(\alpha_\rmR\mathbf{r}_\mathrm{o}(t-\tau)+\beta_\rmR\mathbf{r}_\mathrm{o}^*(t-\tau)\right)}_{\boldsymbol{\mu}(t)}+\underbrace{\left(\alpha_\rmR \mathbf{W}^\rmH\mathbf{n}(t)+\beta_\rmR\mathbf{W}^\rmT \mathbf{n}^*(t)\right)}_{\mathbf{z}(t)}.\label{r_mu_z}
	\end{align}
	In order to simplify the exposition, we assume orthogonal beams such that $\mathbf{W}^\rmH\mathbf{W}=\sigma_b^2\mathbf{I}_{N_\rmB}$, in which $\sigma_b^2$ is the power per beam. This is a reasonable assumption due to the sparse transmission in 5G mmWave channels \cite{overview}. Consequently, the noise variance can be written as
	\begin{subequations}\label{noisevar}
		\begin{align}
		\boldsymbol{\Sigma}_z=\mathbb{E}\left[\mathbf{z}(t)\mathbf{z}^\rmH(t)\right]&=N_0\sigma_b^2\left(|\alpha_\rmR|^2+|\beta_\rmR|^2\right)\mathbf{I}_{N_\rmB}\label{noise1}\\
		&=\underbrace{\frac{1}{2}N_0(1+m_\rmR ^2)\sigma_b^2}_{\triangleq\sigma_z^2}\mathbf{I}_{N_\rmB}\label{noise2}.
		\end{align}
	\end{subequations}
	where \eqref{noise1} follows from the fact that $\mathbb{E}\left[\mathbf{n}(t)\mathbf{n}^\rmT(t)\right]=\mathbf{0}$, and \eqref{noise2} follows from \eqref{rx_iq}. Note that as $\epsilon_\rmR$ increases linearly, the noise covariance at the receiver increases quadratically. 
	
	From \eqref{noisevar}, it is clear that the only parameter in $\boldsymbol{\varphi}_\mathrm{C}$ that $\sigma_z^2$ depends on is $\epsilon_\rmR$. Thus, from \cite{Statistical}, it can be shown that
	\begin{align}\label{jepsilonr}
	{J}_{\epsilon_\rmR\epsilon_\rmR}&=\frac{1}{\sigma_z^2}\int_{0}^{T_0}\mathbb{E}\left\|\frac{\partial\boldsymbol{\mu}(t)}{\partial{\epsilon_\rmR}}\right\| ^2\rmd t+\frac{T_0 N_\rmB^2}{2(\sigma_z^2)^2}\left(\frac{\partial{\sigma_z}^2}{\partial{\epsilon_\rmR}}\right)^2
	\end{align}
	while for all the other parameters in $\boldsymbol{\varphi}_\mathrm{C}$, we have
	\begin{align}
	{J}_{xy}&\triangleq \int_{0}^{T_0}\mathbb{E}\left[\Re\left\{\frac{\partial\boldsymbol{\mu}^\rmH(t)}{\partial{x}}\left(\boldsymbol{\Sigma_Z}\right)^{-1}\frac{\partial\boldsymbol{\mu}(t)}{\partial{y}}\right\}\right] \rmd t,\notag\\
	&=\frac{1}{\sigma_z^2}\int_{0}^{T_0}\mathbb{E}\left[\Re\left\{\frac{\partial\boldsymbol{\mu}^\rmH(t)}{\partial{x}}\frac{\partial\boldsymbol{\mu}(t)}{\partial{y}}\right\}\right] \rmd t,\label{jxyindependent}
	\end{align}
	where $x,y\in\left\{\phi_\rmR,\phi_\rmT,\tau,\gamma_\rmR,\gamma_\rmI,\epsilon_\rmT,\psi_\rmR,\psi_\rmT\right\}$, $T_0\approx N_\rms T_\rms $ is the observation time and $N_\rms $ is the number of pilot symbols. The full derivation of the elements of \eqref{jphiph} is provided in Appendix \ref{appendix}.
	
	The parameters in $\boldsymbol{\varphi}_\mathrm{C}$ can be divided into two groups: geometrical parameters providing information useful for positioning, and nuisance parameters. We are mainly interested in the equivalent FIM \cite{Zohair2017} of the geometrical parameters that accounts for the nuisance parameters. Towards that, defining the vector of geometrical parameters as $\boldsymbol{\varphi}_\mathrm{G}\triangleq [ \phi_\rmR, \phi_\rmT, \tau]^\rmT $, and the vector of nuisance parameters as $\boldsymbol\varphi_\mathrm{N}\triangleq [\gamma_\rmR, \gamma_\rmI, \epsilon_\rmR, \epsilon_\rmT, \psi_\rmR, \psi_\rmT]^\rmT$, we can write \eqref{jphiph} in block form as 
	\begin{align}\label{jphiphd}
	\mathbf{J}_\mathrm{C}=\begin{bmatrix} 
	\mathbf{J}_\mathrm{G}&\mathbf{J}_{\mathrm{GN}}\\
	\mathbf{J}_{\mathrm{GN}}^\mathrm{T}&\mathbf{J}_{\mathrm{N}}
	\end{bmatrix} \;\in\mathbb{R}^{9\times{9}},
	\end{align}
	where $\mathbf{J}_\mathrm{G}\in\mathbb{R}^{3\times{3}}$ and $\mathbf{J}_\mathrm{N}\in\mathbb{R}^{6\times{6}}$ are the FIMs of $\boldsymbol{\varphi}_\mathrm{G}$ and $\boldsymbol{\varphi}_\mathrm{N}$, respectively, while $\mathbf{J}_\mathrm{GN}$ is the mutual information matrix of $\boldsymbol{\varphi}_\mathrm{G}$ and $\boldsymbol{\varphi}_\mathrm{N}$. Consequently, the EFIM of $\boldsymbol{\varphi}_\mathrm{G}$ is computed using Schur complement as \cite{Shen2010_2}
	\begin{align}\label{jech}
	\mathbf{J}_\mathrm{G}^\rme=\mathbf{J}_\mathrm{G}-\mathbf{J}_{\mathrm{GN}}\mathbf{J}^{-1}_{\mathrm{N}}\mathbf{J}^\rmT_{\mathrm{GN}}.
	\end{align}
	Note that the minus sign in \eqref{jech} indicates loss of information due to the nuisance parameters.
	\section{FIM of Location Parameters}
	As highlighted earlier, our goal is to derive the PEB and OEB from the intermediary parameters, i.e., channel parameter. To this end, the FIM of position and orientation, $\boldsymbol{\varphi}_\mathrm{L}\triangleq[p_x,p_y,\phi_0]^\rmT$, can be computed via a transformation of parameters as follows \cite{Statistical}
	\begin{equation}
	\mathbf{J}^\rme_{\mathrm{L}}\triangleq  \boldsymbol\Upsilon \mathbf{J}^\rme_{\mathrm{G}} \boldsymbol\Upsilon^\rmT,
	\end{equation}
	where $\boldsymbol\Upsilon$ is the transformation matrix, given by the Jacobean
	\begin{align}\label{gamma}
	\boldsymbol\Upsilon=\frac{\partial\boldsymbol{\varphi}^\rmT_\mathrm{G}}{\partial\boldsymbol{\varphi}_\mathrm{L}}=\begin{bmatrix}
	\frac{\partial\phi_\mathrm{R}}{\partial p_\mathrm{x}}&\frac{\partial\phi_\mathrm{T}}{\partial p_\mathrm{x}}&\frac{\partial\tau}{\partial p_\mathrm{x}}\\
	\frac{\partial\phi_\mathrm{R}}{\partial p_\mathrm{y}}&\frac{\partial\phi_\mathrm{T}}{\partial p_\mathrm{y}}&\frac{\partial\tau}{\partial p_\mathrm{y}}\\
	\frac{\partial\phi_\mathrm{R}}{\partial\phi_0}&\frac{\partial\phi_\mathrm{T}}{\partial\phi_0}&\frac{\partial\tau}{\partial\phi_0}
	\end{bmatrix}\in\mathbb{R}^{3\times{3}}.
	\end{align}
	The entries of $\boldsymbol\Upsilon$ can be obtained from the relationships between the UE and BS highlighted in the geometry shown in Fig.~\ref{2DUEBS}. That is, defining $c$ as the propagation speed
	\begin{subequations}
		\begin{align}
		\tau&=\frac{\|\mathbf{p}\|}{c},\\
		\phi_\rmR&=\arccos\left(\frac{p_x}{\|\mathbf{p}\|}\right),\\
		\phi_\rmT&=\pi-\phi_0+\arccos\left(\frac{p_x}{\|\mathbf{p}\|}\right).
		\end{align}
	\end{subequations}
	
	Finally, for brevity, define $\mathbf{C}=\left(\mathbf{J}^\rme_{\mathrm{G}}\right)^{-1}$, then the PEB and OEB under IQI can be found as
	\begin{subequations}\label{PEB_OEB}
		\begin{align}
		\text{PEB}_\text{IQ}&=\sqrt{[C]_{1,1}+[C]_{2,2}},\label{SPEB}\\
		\text{OEB}_\text{IQ}&=\sqrt{[C]_{3,3}}.\label{OEB}
		\end{align}
	\end{subequations}
	
	\section{Numerical results}
	\subsection{Simulation Setup}
	We consider a mmWave scenario operating at $f=38$ GHz. The BS is equipped with $N_\rmR=64$ antennas and located at $(0,0)$. On the other hand, the UE is located in a square area, (10 m$\times$10 m), defined by $(p_x,p_y)\in\{(x,y):y\geq|x|\cap y\leq10\sqrt{2}-|x|\}$ and equipped with $N_\rmT=32$ antennas.
	
	We utilize directional beamforming similar to \cite{Zohair2017}, in which beams point toward $\phi_{B,l}, 1\leq l \leq N_\rmB$, such that the transmit and receive beamforming are respectively given by
	\begin{align*}
	\mathbf{f}_l\triangleq \frac{1}{\sqrt{N_\rmB}}\mathbf{a}_\rmT(\phi_{BT,l}),\\
	\mathbf{w}_l\triangleq \frac{1}{\sqrt{N_\rmB}}\mathbf{a}_\rmR(\phi_{BR,l}),
	\end{align*}
	where $\mathbf{a}_\rmT(\phi_{BT,l})$ and $\mathbf{a}_\rmR(\phi_{BR,l})$ have the same structure as \eqref{at}. We chose $N_\rmB=18$, uniformly covering the square area, i.e., $\phi_{BR,l}=\frac{\pi}{4}+\frac{\pi (l-1)}{2(N_\rmB-1)}$.
	
	Furthermore, we assume $\mathbf{s}_\rmT(t)$ to be transmitted through a unit energy ideal $sinc$ pulse shaping filter so that the effective bandwidth, $W_\mathrm{eff}^2=W^2/3$ where $W=125$ MHz. Moreover we use the following parameters $N_0=-170$ dBm/Hz, $N_\rms =16 $, $\sigma_b^2=1$ and $\phi_0=0$. We conduct Monte-Carlo simulations for 120 UE locations and average over 100 iterations to obtain the PEB and OEB degradation due to IQI
	\begin{align*}
	\text{PEB}_\text{deg}&=\frac{\text{PEB}_\text{IQ}-\text{PEB}_\text{match}}{\text{PEB}_\text{match}}\times 100\%,\\
	\text{OEB}_\text{deg}&=\frac{\text{OEB}_\text{IQ}-\text{OEB}_\text{match}}{\text{OEB}_\text{match}}\times 100\%
	\end{align*}
	where PEB$_\text{match}$ and OEB$_\text{match}$ are defined similar to \eqref{PEB_OEB} after dropping $\epsilon_\rmR, \epsilon_\rmT, \psi_\rmR$ and $\psi_\rmT$ from $\boldsymbol{\varphi}_\mathrm{N}$ and setting them to zero in \eqref{stt} and \eqref{r(t)} .  
	\subsection{PEB and OEB with respect to I/Q parameters}
	\begin{figure}[!t]
		\centering
		\includegraphics[scale=0.85]{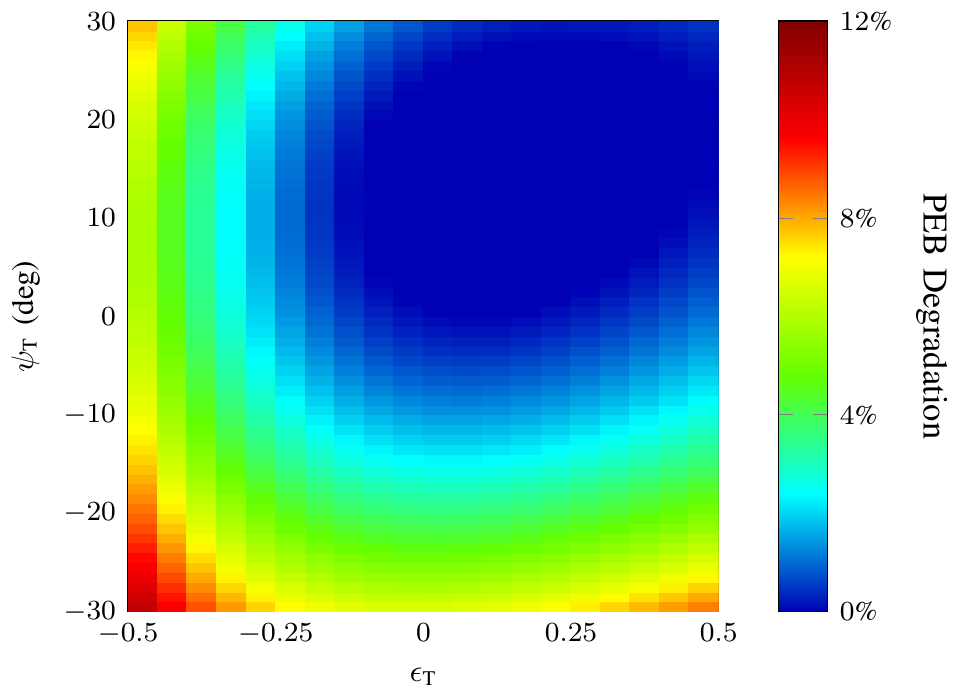}
		\caption{PEB degradation with respect to $\epsilon_\rmT$ and $\psi_\rmT$. $N_\rmT=32$, $N_\rmR=64$,  $N_\rmB=18$ and $\phi_0=0$}
		\label{PEB_T}
	\end{figure}
	\begin{figure}[!t]
	\centering
	\includegraphics[scale=0.85]{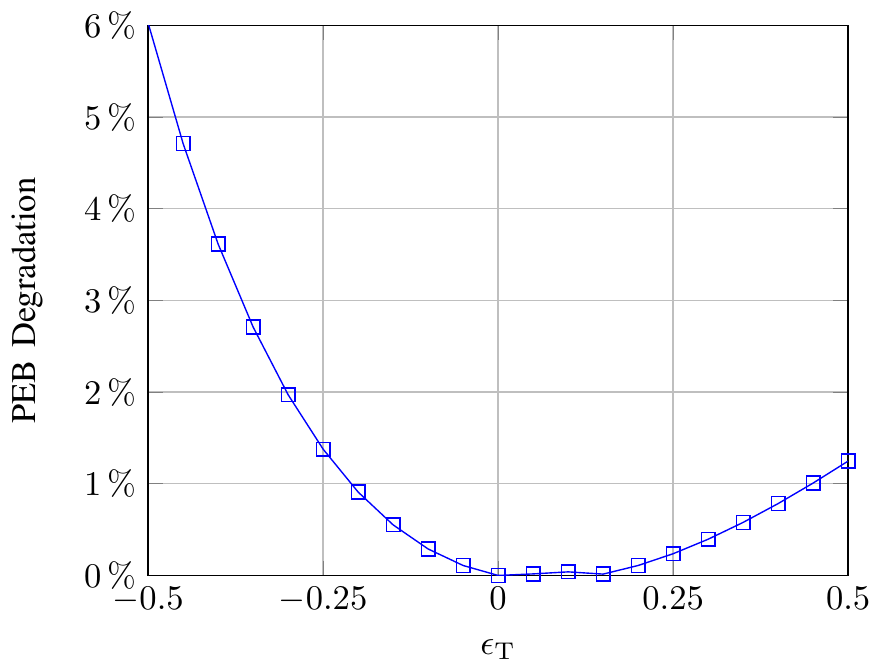}
	\caption{PEB degradation at $\psi_\rmT=0$.}
	\label{fig:PEB_eps}
\end{figure}
	Fig.~\ref{PEB_T} shows user PEB percentage degradation with respect to transmitter I/Q parameters for the considered scenario. For this figure, the receiver parameters are chosen randomly over the ranges $-0.5\leq\epsilon_\rmR\leq0.5$ and $-30^\circ\leq\psi_\rmR\leq30^\circ$. It can be seen that minimum degradation occurs when $\epsilon_\rmT=\psi_\rmT=0$. That is, transmitter I and Q branches are perfectly matched. Moreover, the PEB percentage degradation increases gradually as the imbalance deteriorates by diverging from the point $\epsilon_\rmT=\psi_\rmT=0$. It worth noting that the general behavior of PEB percentage degradation is almost symmetrical along $\psi_\rmT$, unlike $\epsilon_\rmT$. To see this clearer, we present Fig.~\ref{fig:PEB_eps}. It is intuitive that as $\epsilon_\rmT$ increases, the IQI worsens and its impact on PEB degradation increases. However, it can be seen that as $\epsilon_\rmT$ decreases towards negative values, the degradation becomes more pronounced. This occurs because the magnitude of the quadrature carrier signal diminishes, i.e., $m_\rmT \sin(\omega_ct+\psi_\rmT)$ and both $\text{PEB}_\text{IQ}$ and $\text{PEB}_\text{match}$ worsen.
	Considering a system-level evaluation, it can be seen that for relevant values of  $\epsilon_\rmT$ and $\psi_\rmT$, there is up to 15\% bound degradation due to IQI.

	\begin{figure}[!t]
	\centering
	\includegraphics[scale=0.85]{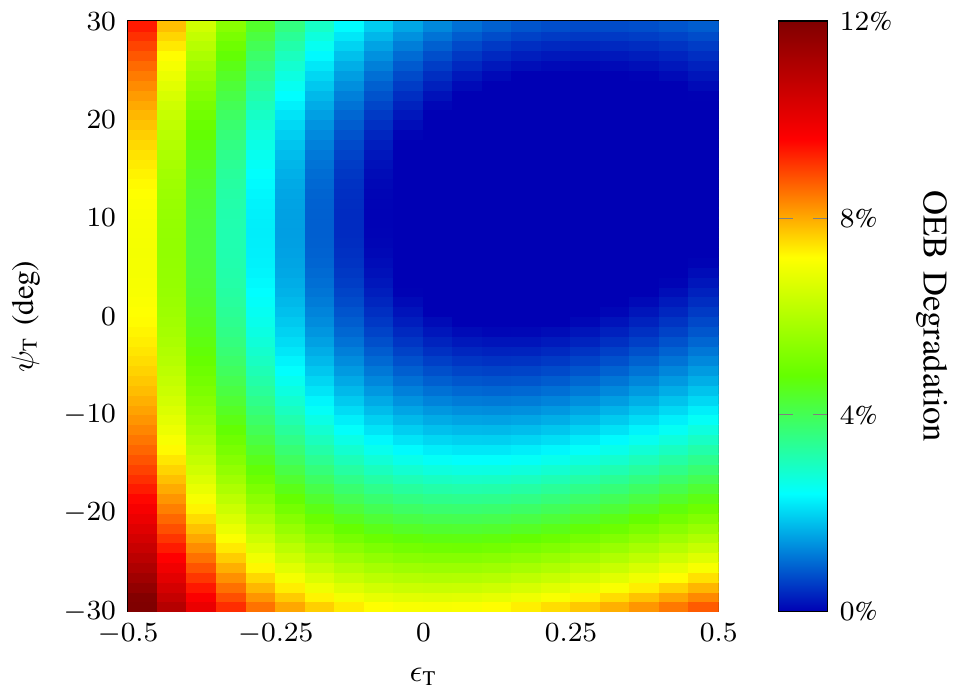}
	\caption{OEB degradation with respect to $\epsilon_\rmT$ and $\psi_\rmT$. $N_\rmT=32$, $N_\rmR=64$,  $N_\rmB=18$ and $\phi_0=0$}
	\label{OEB_T }	
	\end{figure}
	Fig.~\ref{OEB_T } presents the OEB  percentage degradation with respect to the transmitter I/Q parameters. In general, the behavior of OEB degradation is similar to that of the PEB degradation although around the corners OEB is slightly higher. In \cite{Zohair2017}, it has been shown that PEB is a function of DOD and TOA, while OEB is a function of DOA and DOD. Therefore the slight deterioration of OEB with respect to PEB is due the additional in estimating DOA, arising from IQI.
	
\begin{figure}[!t]
	\centering
	\includegraphics[scale=0.85]{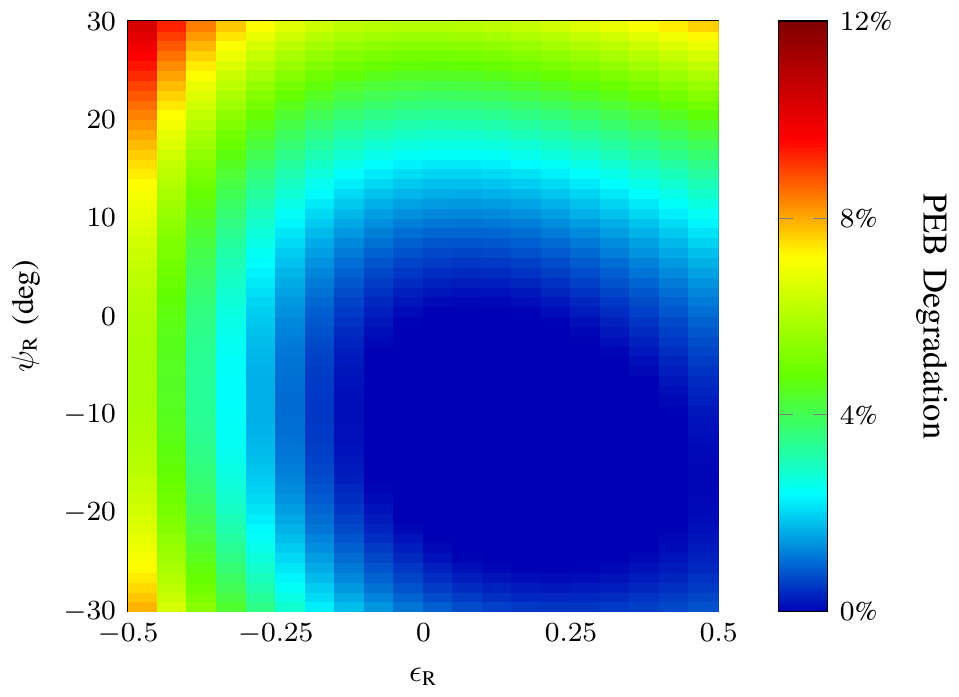}
	\caption{PEB degradation with respect to $\epsilon_\rmR$ and $\psi_\rmR$. $N_\rmT=32$, $N_\rmR=64$,  $N_\rmB=18$ and $\phi_0=0$}
	\label{PEB_R }	
\end{figure}

	Finally, Figs. \ref{PEB_R } and \ref{PEB_R } indicate that PEB and OEB percentage degradation exhibit similar behavior with respect to the receiver I/Q parameters, except that the contour plots are flipped w.r.t $\psi_\rmR=0$. This is attributed the the different signs in \eqref{alpha_beta_t} and \eqref{alpha_beta_r}, resulting in $w_\rmR$ and $w_\rmT$ in \eqref{defs} to have different exponent signs.
	
	\begin{figure}[!t]
	\centering
	\includegraphics[scale=0.85]{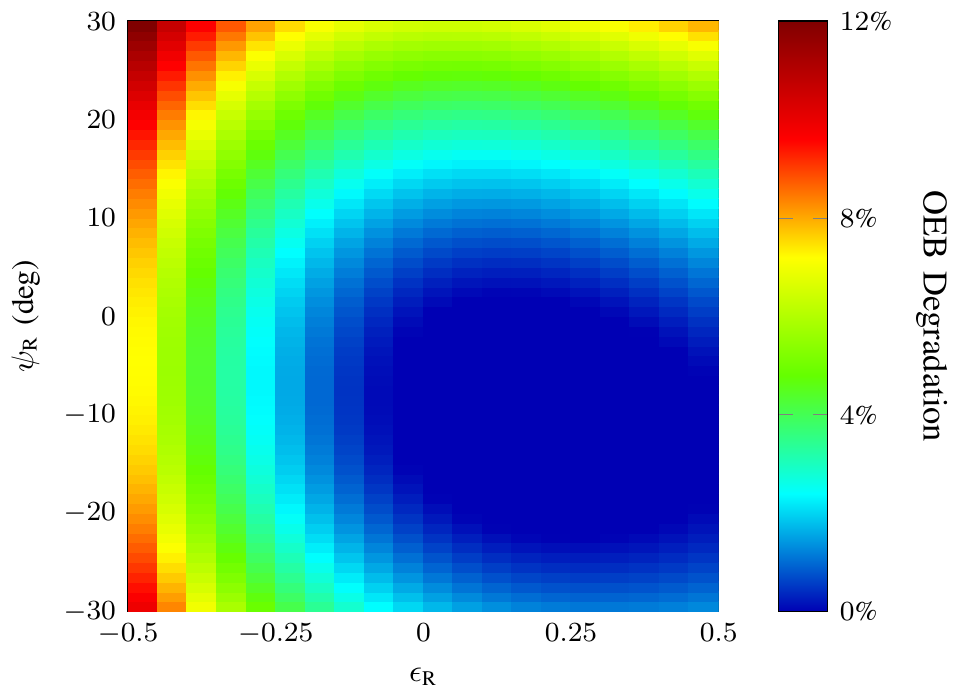}
	\caption{OEB degradation with respect to $\epsilon_\rmR$ and $\psi_\rmR$. $N_\rmT=32$, $N_\rmR=64$,  $N_\rmB=18$ and $\phi_0=0$}
	\label{OEB_R }	
\end{figure}

	\section{Conclusion}
	In this paper, we investigated the effects of I/Q imbalance phenomenon on the position and orientation error bounds. We considered 2D 5G mmWave uplink localization with analog beamforming. Our results show that PEB and OEB degrade by similar amounts with respect to amplitude and phase imbalance. While this degradation is symmetric with respect to the phase imbalance, it is more significant for negative amplitude errors than positive. We also showed that I/Q imbalance can cause up to 12\% increase in the error of location and orientation estimation. For future work, we will consider different transceiver structures such as IF phase-shifting, LO phase shifting and hybrid beamforming.
\appendices
\section{FIM elements}\label{appendix}
To compute \eqref{jepsilonr} and \eqref{jxyindependent}, the following notation is introduced 
\begin{subequations}\label{notation}
\begin{align}
&\mathbf{p}_\rmR\triangleq\frac{\partial\mathbf{r}_{o_2}(t-\tau)}{\partial{\phi_\rmR }}=\gamma\mathbf{W}^\rmH\mathbf{k}_\rmR\mathbf{a}_\rmT^\rmH\mathbf{F}\mathbf{s}_\rmT(t-\tau),\\
&\mathbf{p}_\rmT\triangleq\frac{\partial\mathbf{r}_{o_2}(t-\tau)}{\partial{\phi_\rmT}}=\gamma\mathbf{W}^\rmH\mathbf{a}_\rmR\mathbf{k}_\rmT^\rmH\mathbf{F}\mathbf{s}_\rmT(t-\tau),\\
&\mathbf{k}_\rmR\triangleq\frac{\partial\mathbf{a}_\rmR}{\partial{\phi_\rmR }},\\
&\mathbf{k}_\rmT\triangleq\frac{\partial\mathbf{a}_\rmT}{\partial{\phi_\rmT}},\\
&\dot{\mathbf{r}}_{o_2}(\tau)\triangleq\frac{\partial\mathbf{r}_{o_2}(\tau)}{\partial{\tau}}=\gamma\mathbf{W}^\rmH\mathbf{a}_\rmR\mathbf{a}_\rmT^\rmH\mathbf{F}\frac{\partial\mathbf{s}_\rmT(t-\tau)}{\partial{\tau}},\\
&\mathbf{b}\triangleq\frac{\partial\mathbf{r}_{o_2}(t-\tau)}{\partial{\gamma_\rmR }}=\mathbf{W}^\rmH\mathbf{a}_\rmR\mathbf{a}_\rmT^\rmH\mathbf{F}\mathbf{s}_\rmT(t-\tau),\\
&\mathbf{e}\triangleq\frac{\partial\mathbf{r}_{o_2}(t-\tau)}{\partial{\epsilon_\rmT}}=\gamma\mathbf{W}^\rmH\mathbf{a}_\rmR\mathbf{a}_\rmT^\rmH\mathbf{F}\frac{\partial\mathbf{s}_\rmT(t-\tau)}{\partial{\epsilon_\rmT}},\\ 
&\quad=\frac12e^{j\psi_\rmT}\gamma\mathbf{W}^\rmH\mathbf{a}_\rmR\mathbf{a}_\rmT^\rmH\mathbf{F}(\mathbf{s}(t-\tau)-\mathbf{s}^*(t-\tau)).
\end{align}
\end{subequations}
Note that we drop the angle parameters from $\mathbf{a}_\rmR$ and $\mathbf{a}_\rmT$ for brevity. Subsequently, from \eqref{r_mu_z}, it can be shown that
\begin{subequations}\label{eq:derivatives}
\begin{align}
\frac{\partial\boldsymbol{\mu}(t)}{\partial{\phi_\rmR }}=&\sqrt{E_\rms N_\rmR N_\rmT}(\alpha_\rmR \mathbf{p}_\rmR+\beta_\rmR \mathbf{p}_\rmR^*),\\
\frac{\partial\boldsymbol{\mu}(t)}{\partial{\phi_\rmT}}=&\sqrt{E_\rms N_\rmR N_\rmT}(\alpha_\rmR \mathbf{p}_\rmT+\beta_\rmR \mathbf{p}_\rmT^*),\\
\frac{\partial\boldsymbol{\mu}(t)}{\partial{\tau}}=&\sqrt{E_\rms N_\rmR N_\rmT}(\alpha_\rmR \dot{\mathbf{r}}_{o_2}(t-\tau)+\beta_\rmR \dot{\mathbf{r}}^*_{o_2}(t-\tau)),\\
\frac{\partial\boldsymbol{\mu}(t)}{\partial{\gamma_\rmR }}=&\sqrt{E_\rms N_\rmR N_\rmT}(\alpha_\rmR \mathbf{b}+\beta_\rmR \mathbf{b}^*),\\
\frac{\partial\boldsymbol{\mu}(t)}{\partial{\gamma_I}}=&j\sqrt{E_\rms N_\rmR N_\rmT}(\alpha_\rmR \mathbf{b}-\beta_\rmR \mathbf{b}^*),\\
\frac{\partial\boldsymbol{\mu}(t)}{\partial{\epsilon_\rmR }}=&\frac12\sqrt{E_\rms N_\rmR N_\rmT}(e^{-j\psi_\rmR }\mathbf{r}_{o_2}(t-\tau)-e^{j\psi_\rmR }\mathbf{r}^*_{o_2}(t-\tau)),\\
\frac{\partial\boldsymbol{\mu}(t)}{\partial{\epsilon_\rmT}}=&\sqrt{E_\rms N_\rmR N_\rmT}(\alpha_\rmR \mathbf{e}+\beta_\rmR \mathbf{e}^*),\\
\frac{\partial\boldsymbol{\mu}(t)}{\partial{\psi_\rmR }}=&-\frac{1}{2}jm_\rmR\sqrt{E_\rms N_\rmR N_\rmT}(e^{-j\psi_\rmR }\mathbf{r}_{o_2}(t-\tau)+e^{j\psi_\rmR }\mathbf{r}^*_{o_2}(t-\tau)),\\
\frac{\partial\boldsymbol{\mu}(t)}{\partial{\psi_\rmT}}=&jm_\rmT \sqrt{E_\rms N_\rmR N_\rmT}(\alpha_\rmR \mathbf{e}+\beta_\rmR \mathbf{e}^*)=jm_\rmT \frac{\partial\boldsymbol{\mu}(t)}{\partial{\epsilon_\rmT}},\\
\frac{\partial\boldsymbol{\Sigma_Z}}{\partial{\epsilon_\rmR }}=&\sigma_n^2\sigma_b^2(1+\epsilon_\rmR )\mathbf{I}_{N_\mathrm{B}}.
\end{align}
\end{subequations}
Substituting \eqref{eq:derivatives} in \eqref{jxyindependent} and \eqref{jepsilonr} and using the following relationships, proven in Appendix B 
\begin{subequations}
	\begin{align}
	&\int_{0}^{T_0}\mathbb{E}\left[\mathbf{s}_\rmT(t-\tau)\mathbf{s}^\rmH_\rmT(t-\tau)\right]\rmd t=\frac12(1+m_\rmT^2)N_\rms \mathbf{I}_{N_\mathrm{B}},\\
	&\int_{0}^{T_0}\mathbb{E}\left[\mathbf{s}_\rmT(t-\tau)\mathbf{s}^\rmT _\rmT(t-\tau)\right]\rmd t\triangleq \frac12(1-m_\rmT^2\rme^{j2\psi_\rmT}) N_\rms \mathbf{I}_{N_\mathrm{B}},\\
	&\int_{0}^{T_0}\mathbb{E}\left[\frac{\partial\mathbf{s}_\rmT(t-\tau)}{\partial\tau}\frac{\partial\mathbf{s}^\rmH_\rmT(t-\tau)}{\partial\tau}\right]\rmd t=2\pi^2N_\rms (1+m_\rmT^2) W^2_\mathrm{eff}\mathbf{I}_{N_\mathrm{B}}, \\
	&\int_{0}^{T_0}\mathbb{E}\left[\frac{\partial\mathbf{s}_\rmT(t-\tau)}{\partial\tau}\frac{\partial\mathbf{s}^\rmT _\rmT(t-\tau)}{\partial\tau}\right]\rmd t=4\pi^2 N_\rms (1-m_\rmT^2\rme^{j2\psi_\rmT}) W^2_\mathrm{eff}\mathbf{I}_{N_\mathrm{B}},\\
	&\int_{0}^{T_0}\mathbb{E}\left[\frac{\partial\mathbf{s}_\rmT(t-\tau)}{\partial\tau}\mathbf{s}^\rmH_\rmT(t-\tau)\right]\rmd t=\int_{0}^{T_0}\mathbb{E}\left[\frac{\partial\mathbf{s}_\rmT(t-\tau)}{\partial\tau}\mathbf{s}^\rmT _\rmT(t-\tau)\right]\rmd t=\mathbf{0}.
	\end{align}
\end{subequations}
it can be shown that
\begin{align}
&\boldsymbol{J}_{\phi_\rmR \phi_\rmR }\triangleq\frac{1}{\sigma_z^2}\int_{0}^{T_0}\mathbb{E}\left[\Re\left\{\frac{\partial\boldsymbol{\mu}^\rmH(t)}{\partial{\phi_\rmR }}\frac{\partial\boldsymbol{\mu}(t)}{\partial{\phi_\rmR }}\right\}\right]\rmd t\notag,\\
&=\frac{E_\rms N_\rmR N_\rmT }{\sigma_z^2}\int_{0}^{T_0}\mathbb{E}\left[\Re\left\{(\alpha_\rmR \mathbf{p}_\rmR+\beta_\rmR \mathbf{p}_\rmR^*)^\rmH(\alpha_\rmR \mathbf{p}_\rmR+\beta_\rmR \mathbf{p}_\rmR^*)\right\}\right]\rmd t\notag,\\
&=\frac{E_\rms N_\rmR N_\rmT }{\sigma_z^2}\int_{0}^{T_0}(|\alpha_\rmR |^2+|\beta_\rmR|^2)\mathbb{E}\left[\mathbf{p}_\rmR^\rmH\mathbf{p}_\rmR\right]+2\Re\left\{\alpha_\rmR \beta^*_\rmR\mathbb{E}\left[ \mathbf{p}^\rmT _\rmR \mathbf{p}_\rmR\right]  \rmd t\right\},\\
&=\frac{E_\rms N_\rmR N_\rmT }{2\sigma_z^2}\left((1+m_\rmR^2)\int_{0}^{T_0}\mathbb{E}\left[\mathbf{p}_\rmR^\rmH\mathbf{p}_\rmR\right]\rmd t+\Re\left\{(1-m_\rmR^2\rme^{-j2\psi_\rmR})\int_{0}^{T_0}\mathbb{E}\left[ \mathbf{p}^\rmT _\rmR \mathbf{p}_\rmR\right]  \rmd t\right\}\right)\label{jphrphir}.
\end{align}
where \eqref{jphrphir} is obtained using \eqref{rx_iq}. From \eqref{notation}, it is straight-forward that
\begin{align}\nonumber
\int_{0}^{T_0}\mathbb{E}\left[\mathbf{p}_\rmR^\rmH\mathbf{p}_\rmR\right] \rmd t&=\int_{0}^{T_0}\mathbb{E}\left[(\gamma\mathbf{W}^\rmH\mathbf{k}_\rmR\mathbf{a}_\rmT^\rmH\mathbf{F}\mathbf{s}_\rmT(t-\tau))^\rmH(\gamma\mathbf{W}^\rmH\mathbf{k}_\rmR\mathbf{a}_\rmT^\rmH\mathbf{F}\mathbf{s}_\rmT(t-\tau))\right]\rmd t,\\\nonumber
&=\int_{0}^{T_0}\mathbb{E}\left[(\mathbf{s}^\rmH_\rmT(t-\tau)\mathbf{F}^\rmH\mathbf{a}_\rmT\mathbf{k}_\rmR^\rmH\mathbf{W}\gamma^*)(\gamma\mathbf{W}^\rmH\mathbf{k}_\rmR\mathbf{a}_\rmT^\rmH\mathbf{F}\mathbf{s}_\rmT(t-\tau))\right\}\rmd t,\\\nonumber
&=|\gamma|^2\mathbf{a}_\rmT^\rmH\mathbf{F}\int_{0}^{T_0}\mathbb{E}\left[\mathbf{s}_\rmT(t-\tau)\mathbf{s}^\rmH_\rmT(t-\tau)\right] \rmd t~\mathbf{F}^\rmH\mathbf{a}_\rmT\mathbf{k}_\rmR^\rmH\mathbf{W}\mathbf{W}^\rmH\mathbf{k}_\rmR,\\\nonumber
&=\frac12N_\rms|\gamma|^2(1+m_\rmT^2) (\mathbf{a}_\rmT^\rmH\mathbf{F}\mathbf{F}^\rmH\mathbf{a}_\rmT)(\mathbf{k}_\rmR^\rmH\mathbf{W}\mathbf{W}^\rmH\mathbf{k}_\rmR).
\end{align}
Similarly, it can be shown that 
\begin{align*}
\int_{0}^{T_0}\mathbb{E}\left[ \mathbf{p}^\rmT _\rmR \mathbf{p}_\rmR\right]  \rmd t=\frac12 N_\rms|\gamma|^2(1-m_\rmT^2\rme^{j2\psi_\rmT}) (\mathbf{a}_\rmT^\rmH\mathbf{F}\mathbf{F}^\rmT\mathbf{a}_\rmT^*)(\mathbf{k}_\rmR^\rmT\mathbf{W}^*\mathbf{W}^\rmH\mathbf{k}_\rmR).
\end{align*}
Substituting in \eqref{jphrphir}, we can write,
\begin{align}
\boldsymbol{J}_{\phi_\rmR \phi_\rmR }=&\frac{E_\rms N_\rmR N_\rmT N_\rms  }{4\sigma_z^2}\Big(|\gamma|^2(1+m_\rmT^2)(1+m_\rmR^2) (\mathbf{a}_\rmT^\rmH\mathbf{F}\mathbf{F}^\rmH\mathbf{a}_\rmT)(\mathbf{k}_\rmR^\rmH\mathbf{W}\mathbf{W}^\rmH\mathbf{k}_\rmR)\notag\\
&+\Re\left\{(1-m_\rmT^2\rme^{j2\psi_\rmT})(1-m_\rmR^2\rme^{-j2\psi_\rmR})\gamma^2(\mathbf{a}_\rmT^\rmH\mathbf{F}\mathbf{F}^\rmT\mathbf{a}_\rmT^*)(\mathbf{k}_\rmR^\rmT \mathbf{W}^*\mathbf{W}^\rmH\mathbf{k}_\rmR)\right\}\Big).\label{jphirphirbeforefinal}
\end{align}
Defining the following notation
\begin{subequations}\label{defs}
\begin{align}
\eta\triangleq&\frac{E_\rms N_\rmR N_\rmT N_\rms}{4\sigma_z^2},&\gamma\triangleq&|\gamma|\rme^{j\theta},\\
g_\rmR\triangleq&(1+m_\rmR^2), &g_\rmT\triangleq&(1+m_\rmT^2),\\
w_\rmR\triangleq&(1-m_\rmR^2\rme^{-j2\psi_\rmR}),&w_\rmT\triangleq&(1-m_\rmT^2\rme^{j2\psi_\rmT}).
\end{align}
\end{subequations}
and following the same procedure outlined above the entries of the FIM in \eqref{jphiph} can be shown to be  
\begin{align*} 
{J}_{\phi_\rmR\phi_\rmR}&=\eta|\gamma|^2\big(g_\rmR g_\rmT\|\mathbf{a}_\rmT^\rmH\mathbf{F}\|^2\|\mathbf{k}_\rmR^\rmH\mathbf{W}\|^2+\Re\{w_\rmT w_\rmR \rme^{j2\theta}(\mathbf{a}_\rmT^\rmH\mathbf{F}\mathbf{F}^\rmT\mathbf{a}_\rmT^*)(\mathbf{k}_\rmR^\rmT\mathbf{W}^*\mathbf{W}^\rmH\mathbf{k}_\rmR)\}\big)
,\\{J}_{\phi_\rmT\phi_\rmT}&=\eta|\gamma|^2\big( g_\rmR g_\rmT\|\mathbf{k}_\rmT^\rmH\mathbf{F}\|^2\|\mathbf{a}_\rmR^\rmH\mathbf{W}\|^2
+\Re\{w_\rmT w_\rmR \rme^{j2\theta}(\mathbf{k}_\rmT^\rmH\mathbf{F}\mathbf{F}^\rmT\mathbf{k}_\rmT^*)(\mathbf{a}_\rmR^\rmT\mathbf{W}^*\mathbf{W}^\rmH\mathbf{a}_\rmR)\}\big)
,\\{J}_{\gamma_\rmR \gamma_\rmR }&=\eta\big(g_\rmR g_\rmT\|\mathbf{a}_\rmT^\rmH\mathbf{F}\|^2\|\mathbf{a}_\rmR^\rmH\mathbf{W}\|^2
+\Re\{w_\rmT w_\rmR (\mathbf{a}_\rmT^\rmH\mathbf{F}\mathbf{F}^\rmT\mathbf{a}_\rmT^*)(\mathbf{a}_\rmR^\rmT\mathbf{W}^*\mathbf{W}^\rmH\mathbf{a}_\rmR)\}\big)
,\\{J}_{\gamma_I\gamma_I}&=\eta\big(g_\rmR g_\rmT\|\mathbf{a}_\rmT^\rmH\mathbf{F}\|^2\|\mathbf{a}_\rmR^\rmH\mathbf{W}\|^2
-\Re\{w_\rmT w_\rmR (\mathbf{a}_\rmT^\rmH\mathbf{F}\mathbf{F}^\rmT\mathbf{a}_\rmT^*)(\mathbf{a}_\rmR^\rmT\mathbf{W}^*\mathbf{W}^\rmH\mathbf{a}_\rmR)\}\big)
,\\{J}_{\tau\tau}&=4\pi^2W^2_\mathrm{eff}\eta|\gamma|^2\big( g_\rmR g_\rmT\|\mathbf{a}_\rmT^\rmH\mathbf{F}\|^2\|\mathbf{a}_\rmR^\rmH\mathbf{W}\|^2
+\Re\{w_\rmR w_\rmT\rme^{j2\theta} (\mathbf{a}_\rmT^\rmH\mathbf{F}\mathbf{F}^\rmT\mathbf{a}_\rmT^*)(\mathbf{a}_\rmR^\rmT\mathbf{W}^*\mathbf{W}^\rmH\mathbf{a}_\rmR)\}\big)
,\\{J}_{\epsilon_\rmR \epsilon_\rmR }&=2m_\rmR^2N_B^2T_0/g_\rmR^2+\eta|\gamma|^2\big(g_\rmT\|\mathbf{a}_\rmT^\rmH\mathbf{F}\|^2\|\mathbf{a}_\rmR^\rmH\mathbf{W}\|^2
-\Re\{w_\rmT e^{-2j\psi_\rmR } \rme^{j2\theta}(\mathbf{a}_\rmT^\rmH\mathbf{F}\mathbf{F}^\rmT\mathbf{a}^*_\rmT)(\mathbf{a}_\rmR^\rmT\mathbf{W}^*\mathbf{W}^\rmH\mathbf{a}_\rmR)\}\big)
,\\{J}_{\epsilon_\rmT\epsilon_\rmT}&=\eta|\gamma|^2 m_\rmT^2\cos^2(\psi_\rmT)\big(g_\rmR\|\mathbf{a}_\rmT^\rmH\mathbf{F}\|^2\|\mathbf{a}_\rmR^\rmH\mathbf{W}\|^2 
-\Re\{w_\rmR e^{2j\psi_\rmT} \rme^{j2\theta}(\mathbf{a}_\rmT^\rmH\mathbf{F}\mathbf{F}^\rmT\mathbf{a}^*_\rmT)(\mathbf{a}_\rmR^\rmT\mathbf{W}^*\mathbf{W}^\rmH\mathbf{a}_\rmR)\}\big)
,\\{J}_{\psi_\rmR \psi_\rmR }&=\eta|\gamma|^2 m_\rmR^2\big( g_\rmT\|\mathbf{a}_\rmT^\rmH\mathbf{F}\|^2\|\mathbf{a}_\rmR^\rmH\mathbf{W}\|^2
+\Re\{w_\rmT e^{-2j\psi_\rmR } \rme^{j2\theta}(\mathbf{a}_\rmT^\rmH\mathbf{F}\mathbf{F}^\rmT\mathbf{a}^*_\rmT)(\mathbf{a}_\rmR^\rmT\mathbf{W}^*\mathbf{W}^\rmH\mathbf{a}_\rmR)\} \big)
,\\{J}_{\psi_\rmT\psi_\rmT}&=\eta|\gamma|^2 m_\rmT^4\cos^2(\psi_\rmT)\big(g_\rmR\|\mathbf{a}_\rmT^\rmH\mathbf{F}\|^2\|\mathbf{a}_\rmR^\rmH\mathbf{W}\|^2 
-\Re\{w_\rmR e^{2j\psi_\rmT} \rme^{j2\theta}(\mathbf{a}_\rmT^\rmH\mathbf{F}\mathbf{F}^\rmT\mathbf{a}^*_\rmT)(\mathbf{a}_\rmR^\rmT\mathbf{W}^*\mathbf{W}^\rmH\mathbf{a}_\rmR)\}\big)
,\\{J}_{\phi_\rmR \phi_\rmT}&=\eta|\gamma|^2\big( g_\rmT g_\rmR\Re\{(\mathbf{k}_\rmT^\rmH\mathbf{F}\mathbf{F}^\rmH\mathbf{a}_\rmT)(\mathbf{k}_\rmR^\rmH\mathbf{W}\mathbf{W}^\rmH\mathbf{a}_\rmR)\} 
+\Re\{w_\rmT w_\rmR \rme^{j2\theta}(\mathbf{k}_\rmT^\rmH\mathbf{F}\mathbf{F}^\rmT\mathbf{a}_\rmT^*)(\mathbf{k}_\rmR^\rmT\mathbf{W}^*\mathbf{W}^\rmH\mathbf{a}_\rmR)\}\big)
,\\{J}_{\phi_\rmR \gamma_\rmR }&=\eta\big(g_\rmT g_\rmR\|\mathbf{a}_\rmT^\rmH\mathbf{F}\|^2\Re\{\gamma^*(\mathbf{k}_\rmR^\rmH\mathbf{W}\mathbf{W}^\rmH\mathbf{a}_\rmR)\}
+\Re\{w_\rmT w_\rmR \gamma(\mathbf{a}_\rmT^\rmH\mathbf{F}\mathbf{F}^\rmT\mathbf{a}_\rmT^*)(\mathbf{k}_\rmR^\rmT\mathbf{W}^*\mathbf{W}^\rmH\mathbf{a}_\rmR)\}\big)
,\\{J}_{\phi_\rmR \gamma_I}&=-\eta\big(2 g_\rmT m_\rmR\cos(\psi_\rmR)\|\mathbf{a}_\rmT^\rmH\mathbf{F}\|^2\Im\{\gamma^*(\mathbf{k}_\rmR^\rmH\mathbf{W}\mathbf{W}^\rmH\mathbf{a}_\rmR\} 
+\Im\{w_\rmT w_\rmR \gamma(\mathbf{a}_\rmT^\rmH\mathbf{F}\mathbf{F}^\rmT\mathbf{a}_\rmT^*)(\mathbf{k}_\rmR^\rmT\mathbf{W}^*\mathbf{W}^\rmH\mathbf{a}_\rmR)\}\big)
,\\{J}_{\phi_\rmR \epsilon_\rmR }&=\eta|\gamma|^2 m_\rmR\big(|\gamma|^2( 1+m_\rmT^2)\|\mathbf{a}_\rmT^\rmH\mathbf{F}\|^2\Re\{\mathbf{k}_\rmR^\rmT\mathbf{W}^*\mathbf{W}^\rmT\mathbf{a}_\rmR^*\} 
-\Re\{w_\rmT e^{-2j\psi_\rmR } \rme^{j2\theta}(\mathbf{a}_\rmT^\rmH\mathbf{F}\mathbf{F}^\rmT\mathbf{a}_\rmT^*)(\mathbf{k}_\rmR^\rmT\mathbf{W}^*\mathbf{W}^\rmH\mathbf{a}_\rmR)\}\big)
,\\{J}_{\phi_{R}\epsilon_\rmT}&=\eta|\gamma|^2 m_\rmT^2\cos(\psi_\rmT)\big(g_\rmR\|\mathbf{a}_\rmT^\rmH\mathbf{F}\|^2\Re\{\mathbf{k}_\rmR^\rmH\mathbf{W}\mathbf{W}^\rmH\mathbf{a}_\rmR\} 
-\Re\{w_\rmR \rme^{j2\theta}\rme^{2j\psi_\rmT}(\mathbf{a}_\rmT^\rmH\mathbf{F}\mathbf{F}^\rmT\mathbf{a}_\rmT^*)(\mathbf{k}_\rmR^\rmT\mathbf{W}^*\mathbf{W}^\rmH\mathbf{a}_\rmR)\}\big)
,\\{J}_{\phi_\rmR \psi_\rmR }&=-\eta|\gamma|^2 m_\rmR^2\big(g_\rmT\|\mathbf{a}_\rmT^\rmH\mathbf{F}\|^2\Im\{\mathbf{k}_\rmR^\rmT\mathbf{W}^*\mathbf{W}^\rmT\mathbf{a}_\rmR^*\} 
+\Im\{w_\rmT e^{-2j\psi_\rmR }\rme^{j2\theta}(\mathbf{a}_\rmT^\rmH\mathbf{F}\mathbf{F}^\rmT\mathbf{a}_\rmT^*)(\mathbf{k}_\rmR^\rmT\mathbf{W}^*\mathbf{W}^\rmH\mathbf{a}_\rmR)\}\big)
,\\{J}_{\phi_{R}\psi_\rmT}&=-\eta|\gamma|^2 m_\rmT^3m_\rmR\cos(\psi_\rmT)\cos(\psi_\rmR)\|\mathbf{a}_\rmT^\rmH\mathbf{F}\|^2\Im\{\mathbf{k}_\rmR^\rmH\mathbf{W}\mathbf{W}^\rmH\mathbf{a}_\rmR\}
,\\{J}_{\phi_\rmT\gamma_\rmR }&=\eta\big( g_\rmT g_\rmR\|\mathbf{a}_\rmR^\rmH\mathbf{W}\|^2\Re\{\gamma^*(\mathbf{a}_\rmT^\rmH\mathbf{F}\mathbf{F}^\rmH\mathbf{k}_\rmT)\} 
+\Re\{w_\rmT w_\rmR \gamma(\mathbf{a}_\rmT^\rmH\mathbf{F}\mathbf{F}^\rmT\mathbf{k}_\rmT^*)(\mathbf{a}_\rmR^\rmT\mathbf{W}^*\mathbf{W}^\rmH\mathbf{a}_\rmR)\}\big)
,\\{J}_{\phi_\rmT\gamma_I}&=-\eta\big(g_\rmT g_\rmR\|\mathbf{a}_\rmR^\rmH\mathbf{W}\|^2\Im\{\gamma^*(\mathbf{a}_\rmT^\rmH\mathbf{F}\mathbf{F}^\rmH\mathbf{k}_\rmT)\} 
+\Im\{w_\rmT w_\rmR \gamma(\mathbf{a}_\rmT^\rmH\mathbf{F}\mathbf{F}^\rmT\mathbf{k}_\rmT^*)(\mathbf{a}_\rmR^\rmT\mathbf{W}^*\mathbf{W}^\rmH\mathbf{a}_\rmR)\}\big)
,\\{J}_{\phi_\rmT\epsilon_\rmR }&=\eta|\gamma|^2 m_\rmR\big(g_\rmT\|\mathbf{a}_\rmR^\rmH\mathbf{W}\|^2\Re\{\mathbf{a}_\rmT^\rmT\mathbf{F}^*\mathbf{F}^\rmT\mathbf{k}_\rmT^*\} 
-\Re\{w_\rmT e^{-2j\psi_\rmR}\rme^{j2\theta}(\mathbf{a}_\rmT^\rmH\mathbf{F}\mathbf{F}^\rmT\mathbf{k}_\rmT^*)(\mathbf{a}_\rmR^\rmT\mathbf{W}^*\mathbf{W}^\rmH\mathbf{a}_\rmR)\}\big)
,\\{J}_{\phi_{T}\epsilon_\rmT}&=\eta|\gamma|^2m_\rmT^2\cos(\psi_\rmT)\big(g_\rmR\|\mathbf{a}_\rmR^\rmH\mathbf{W}\|^2\Re\{\mathbf{a}_\rmT^\rmH\mathbf{F}\mathbf{F}^\rmH\mathbf{k}_\rmT\} 
-\Re\{w_\rmR\rme^{j2\theta}\rme^{2j\psi_\rmT}(\mathbf{a}_\rmT^\rmH\mathbf{F}\mathbf{F}^\rmT\mathbf{k}_\rmT^*)(\mathbf{a}_\rmR^\rmT\mathbf{W}^*\mathbf{W}^\rmH\mathbf{a}_\rmR)\}\big)
,\\{J}_{\phi_\rmT\psi_\rmR }&=-\eta|\gamma|^2 m_\rmR^2\big( g_\rmT\|\mathbf{a}_\rmR^\rmH\mathbf{W}\|^2\Im\{\mathbf{a}_\rmT^\rmT\mathbf{F}^*\mathbf{F}^\rmT\mathbf{k}_\rmT^*\} 
+\Im\{w_\rmT e^{-2j\psi_\rmR } \rme^{j2\theta}(\mathbf{a}_\rmT^\rmH\mathbf{F}\mathbf{F}^\rmT\mathbf{k}_\rmT^*)(\mathbf{a}_\rmR^\rmT\mathbf{W}^*\mathbf{W}^\rmH\mathbf{a}_\rmR)\}\big)
,\\{J}_{\phi_{T}\psi_\rmT}&=-2\eta|\gamma|^2 m_\rmR m_\rmT^3\cos(\psi_\rmT)\cos(\psi_\rmR)\|\mathbf{a}_\rmR^\rmH\mathbf{W}\|^2\Im\{\mathbf{a}_\rmT^\rmH\mathbf{F}\mathbf{F}^\rmH\mathbf{k}_\rmT\} 
,\\{J}_{\gamma_\rmR \gamma_I}&=-\eta\Im\{w_\rmT w_\rmR (\mathbf{a}_\rmT^\rmH\mathbf{F}\mathbf{F}^\rmT\mathbf{a}_\rmT^*)(\mathbf{a}_\rmR^\rmT\mathbf{W}^*\mathbf{W}^\rmH\mathbf{a}_\rmR)\}
,\\{J}_{\gamma_\rmR \epsilon_\rmR }&=\eta m_\rmR\big(g_\rmT\gamma_\rmR \|\mathbf{a}_\rmT^\rmH\mathbf{F}\|^2\|\mathbf{a}_\rmR^\rmH\mathbf{W}\|^2-\Re\{w_\rmT\gamma e^{-2j\psi_\rmR}(\mathbf{a}_\rmT^\rmH\mathbf{F}\mathbf{F}^\rmT\mathbf{a}_\rmT^*)(\mathbf{a}_\rmR^\rmT\mathbf{W}^*\mathbf{W}^\rmH\mathbf{a}_\rmR)\}\big)
,\\{J}_{\gamma_{R}\epsilon_\rmT}&=\eta m_\rmT^2\cos(\psi_\rmT)\big(g_\rmR\gamma_\rmR \|\mathbf{a}_\rmT^\rmH\mathbf{F}\|^2\|\mathbf{a}_\rmR^\rmH\mathbf{W}\|^2
-\Re\{w_\rmR \gamma\rme^{2j\psi_\rmT}(\mathbf{a}_\rmT^\rmH\mathbf{F}\mathbf{F}^\rmT\mathbf{a}_\rmT^*)(\mathbf{a}_\rmR^\rmT\mathbf{W}^*\mathbf{W}^\rmH\mathbf{a}_\rmR)\}\big)
,\\{J}_{\gamma_\rmR \psi_\rmR }&=\eta m_\rmR^2\big(g_\rmT\gamma_\rmI\|\mathbf{a}_\rmT^\rmH\mathbf{F}\|^2\|\mathbf{a}_\rmR^\rmH\mathbf{W}\|^2-\Im\{w_\rmT\gamma e^{-2j\psi_\rmR}(\mathbf{a}_\rmT^\rmH\mathbf{F}\mathbf{F}^\rmT\mathbf{a}_\rmT^*)(\mathbf{a}_\rmR^\rmT\mathbf{W}^*\mathbf{W}^\rmH\mathbf{a}_\rmR\}\big)
,\\{J}_{\gamma_{R}\psi_\rmT}&=2\eta m_\rmT^3m_\rmR\gamma_\rmI \cos(\psi_\rmR)\cos(\psi_\rmT)\|\mathbf{a}_\rmT^\rmH\mathbf{F}\|^2\|\mathbf{a}_\rmR^\rmH\mathbf{W}\|^2
,\\{J}_{\gamma_I\epsilon_\rmR }&=\eta m_\rmR\big(\gamma_\rmI  g_\rmT\|\mathbf{a}_\rmT^\rmH\mathbf{F}\|^2\|\mathbf{a}_\rmR^\rmH\mathbf{W}\|^2
+\Im\{w_\rmT \gamma e^{-2j\psi_\rmR}(\mathbf{a}_\rmT^\rmH\mathbf{F}\mathbf{F}^\rmT\mathbf{a}_\rmT^*)(\mathbf{a}_\rmR^\rmT\mathbf{W}^*\mathbf{W}^\rmH\mathbf{a}_\rmR)\}\big)
,\\{J}_{\gamma_{I}\epsilon_\rmT}&= \eta m_\rmT^2\cos(\psi_\rmT)\big(g_\rmR\gamma_\rmI \|\mathbf{a}_\rmT^\rmH\mathbf{F}\|^2\|\mathbf{a}_\rmR^\rmH\mathbf{W}\|^2+\Im\{w_\rmR \gamma \rme^{2j\psi_\rmT}(\mathbf{a}_\rmT^\rmH\mathbf{F}\mathbf{F}^\rmT\mathbf{a}_\rmT^*)(\mathbf{a}_\rmR^\rmT\mathbf{W}^*\mathbf{W}^\rmH\mathbf{a}_\rmR)\}\big)
,\\{J}_{\gamma_I\psi_\rmR }&=-\eta m_\rmR^2\big(\gamma_\rmR g_\rmT\|\mathbf{a}_\rmT^\rmH\mathbf{F}\|^2\|\mathbf{a}_\rmR^\rmH\mathbf{W}\|^2 
+\Re\{w_\rmT \gamma e^{-2j\psi_\rmR }(\mathbf{a}_\rmT^\rmH\mathbf{F}\mathbf{F}^\rmT\mathbf{a}_\rmT^*)(\mathbf{a}_\rmR^\rmT\mathbf{W}^*\mathbf{W}^\rmH\mathbf{a}_\rmR)\}\big)
,\\{J}_{\gamma_{I}\psi_\rmT}&=-\eta m_\rmT^3\cos(\psi_\rmT)\big(g_\rmR\gamma_\rmI\|\mathbf{a}_\rmT^\rmH\mathbf{F}\|^2\|\mathbf{a}_\rmR^\rmH\mathbf{W}\|^2+\Im\{w_\rmR\gamma\rme^{2j\psi_\rmT}(\mathbf{a}_\rmT^\rmH\mathbf{F}\mathbf{F}^\rmT\mathbf{a}_\rmT^*)(\mathbf{a}_\rmR^\rmT\mathbf{W}^*\mathbf{W}^\rmH\mathbf{a}_\rmR)\}\big)
,\\{J}_{\epsilon_\rmR \epsilon_\rmT}&=\eta |\gamma|^2m_\rmR m_\rmT^2\cos(\psi_\rmT)\big(\|\mathbf{a}_\rmT^\rmH\mathbf{F}\|^2\|\mathbf{a}_\rmR^\rmH\mathbf{W}\|^2+\Re\{e^{-2j(\psi_\rmR-\psi_\rmT)}\rme^{j2\theta}(\mathbf{a}_\rmT^\rmH\mathbf{F}\mathbf{F}^\rmT\mathbf{a}^*_\rmT)(\mathbf{a}_\rmR^\rmT\mathbf{W}^*\mathbf{W}^\rmH\mathbf{a}_\rmR)\}\big)
,\\{J}_{\epsilon_\rmR \psi_\rmR }&=-\eta|\gamma|^2 m_\rmR\Im\{w_\rmT e^{-2j\psi_\rmR }\rme^{j2\theta}(\mathbf{a}_\rmT^\rmH\mathbf{F}\mathbf{F}^\rmT\mathbf{a}^*_\rmT)(\mathbf{a}_\rmR^\rmT\mathbf{W}^*\mathbf{W}^\rmH\mathbf{a}_\rmR)\}
,\\{J}_{\epsilon_\rmR \psi_\rmT}&=-\eta|\gamma|^2 m_\rmT^3\cos(\psi_\rmT)\big(\sin(\psi_\rmR)\|\mathbf{a}_\rmT^\rmH\mathbf{F}\|^2\|\mathbf{a}_\rmR^\rmH\mathbf{W}\|^2 
+\Im\{e^{-j(\psi_\rmR-2\psi_\rmT)}\rme^{j2\theta}(\mathbf{a}_\rmT^\rmH\mathbf{F}\mathbf{F}^\rmT\mathbf{a}^*_\rmT)(\mathbf{a}_\rmR^\rmT\mathbf{W}^*\mathbf{W}^\rmH\mathbf{a}_\rmR)\}\big)
,\\{J}_{\epsilon_\rmT\psi_\rmR }&=\eta|\gamma|^2 m_\rmR^2 m_\rmT^2\cos(\psi_\rmT)\Im\{e^{-2j(\psi_\rmR -\psi_\rmT)}\rme^{j2\theta}(\mathbf{a}_\rmT^\rmH\mathbf{F}\mathbf{F}^\rmT\mathbf{a}^*_\rmT)(\mathbf{a}_\rmR^\rmT\mathbf{W}^*\mathbf{W}^\rmH\mathbf{a}_\rmR)\}
,\\{J}_{\psi_\rmT\psi_\rmR }&=\eta|\gamma|^2 m_\rmR^2 m_\rmT^3\cos(\psi_\rmT)\Im\{e^{-2j(\psi_\rmR-\psi_\rmT)}\rme^{j2\theta}(\mathbf{a}_\rmT^\rmH\mathbf{F}\mathbf{F}^\rmT\mathbf{a}^*_\rmT)(\mathbf{a}_\rmR^\rmT\mathbf{W}^*\mathbf{W}^\rmH\mathbf{a}_\rmR)\}.
\end{align*}
\section{Derivation of Signal Correlation}
We start by noting that from \cite{Zohair2017},
\begin{align*}
&\int_{0}^{T_0}\mathbb{E}\left[\mathbf{s}(t-\tau)\mathbf{s}^\rmH(t-\tau)\right]\rmd t=N_\rms \mathbf{I}_{N_\mathrm{B}},\\
&\int_{0}^{T_0}\mathbb{E}\left[\frac{\partial\mathbf{s}(t-\tau)}{\partial\tau}\mathbf{s}^\rmH(t-\tau)\right]\rmd t=\mathbf{0},\\
&\int_{0}^{T_0}\mathbb{E}\left[\frac{\partial\mathbf{s}(t-\tau)}{\partial\tau}\frac{\partial\mathbf{s}^\rmH(t-\tau)}{\partial\tau}\right]\rmd t=4\pi^2N_\rms W_\mathrm{eff}\mathbf{I}_{N_\mathrm{B}},
\end{align*}
where $W_\mathrm{eff}\triangleq\int_{-W/2}^{W/2}f^2| P(f)|^2 \rmd f$ and $W$ is the bandwidth, we can write
\begin{align*}
\int_{0}^{T_0}\mathbb{E}\left[\mathbf{s}_\rmT(t-\tau)\mathbf{s}^\rmH_\rmT(t-\tau)\right]\rmd t&=\int_{0}^{T_0}\mathbb{E}\left[(\alpha_\rmT \mathbf{s}(t)+\beta_\rmT \mathbf{s}^* (t))(\alpha_\rmT^* \mathbf{s}^\rmH(t)+\beta_\rmT^* \mathbf{s}^\rmT (t))\right]\rmd t,\\
&=\int_{0}^{T_0}\mathbb{E}\left[\left(|\alpha_\rmT|^2+|\beta_\rmT|^2\right)\mathbf{s}(t)\mathbf{s}^\rmH(t)+2\Re\{\alpha_\rmT\beta_\rmT^*\mathbf{s}(t)\mathbf{s}^\rmT(t)\}\right]\rmd t,\\
&=(|\alpha_\rmT|^2+|\beta_\rmT|^2)N_\rms \mathbf{I}_{N_\mathrm{B}}=\frac12(1+m_\rmT^2)N_\rms \mathbf{I}_{N_\mathrm{B}}.
\end{align*}
where we used \eqref{alpha_beta_t} and the fact that $\mathbb{E}\left[\mathbf{s}(t)\mathbf{s}^\rmT(t)\right]=\mathbf{0}$, based on the assumption that the real and imaginary parts of $\mathbf{s}(t)$ are independent with zero mean. Similarly, it can be shown that
\begin{align*}
\int_{0}^{T_0}\mathbf{s}_\rmT(t-\tau)\mathbf{s}^\rmT _\rmT(t-\tau)\rmd t&=\int_{0}^{T_0}\mathbb{E}\left[(\alpha_\rmT^2+\beta_\rmT^2)\mathbf{s}(t)\mathbf{s}^\rmT(t)+2\alpha_\rmT\beta_\rmT\mathbf{s}(t)\mathbf{s}^\rmH(t)\right]\rmd t,\\
&=2\alpha_\rmT\beta_\rmT N_\rms \mathbf{I}_{N_\mathrm{B}}=\frac12(1-m_\rmT^2\rme^{j2\psi_\rmT})N_\rms \mathbf{I}_{N_\mathrm{B}}.
\end{align*}
\begin{align*}
\int_{0}^{T_0}\mathbb{E}\left[\frac{\partial\mathbf{s}_\rmT(t-\tau)}{\partial\tau}\frac{\partial\mathbf{s}^\rmH_\rmT(t-\tau)}{\partial\tau}\right]\rmd t&=\int_{0}^{T_0}\mathbb{E}\left[\frac{\partial(\alpha_\rmT \mathbf{s}(t)+\beta_\rmT \mathbf{s}^* (t))}{\partial t}\frac{\partial(\alpha_\rmT^* \mathbf{s}^\rmH(t)+\beta_\rmT^* \mathbf{s}^\rmT (t))}{\partial t}\right] \rmd t,\\
&=\int_{0}^{T_0}\mathbb{E}\left[(|\alpha_\rmT|^2+|\beta_\rmT|^2)\frac{\partial\mathbf{s}(t)}{\partial t}\frac{\partial\mathbf{s}^\rmH(t)}{\partial t}+2\Re\left\{\alpha_\rmT\beta_\rmT^*\frac{\partial\mathbf{s}(t)}{\partial t}\frac{\partial\mathbf{s}^\rmT(t)}{\partial t}\right\}\right]\rmd t,\\ 
&=4\pi^2N_\rms (|\alpha_\rmT|^2+|\beta_\rmT|^2)W^2_\mathrm{eff}\mathbf{I}_{N_\mathrm{B}}=2\pi^2N_\rms (1+m_\rmT^2) W^2_\mathrm{eff}\mathbf{I}_{N_\mathrm{B}}.
\end{align*}
Moreover,
\begin{align*}
\int_{0}^{T_0}\mathbb{E}\left[\frac{\partial\mathbf{s}_\rmT(t-\tau)}{\partial\tau}\frac{\partial\mathbf{s}^\rmT _\rmT(t-\tau)}{\partial\tau}\right]\rmd t&=\int_{0}^{T_0}\mathbb{E}\left[\frac{\partial(\alpha_\rmT \mathbf{s}(t)+\beta_\rmT \mathbf{s}^* (t))}{\partial t}\frac{\partial(\alpha_\rmT \mathbf{s}^\rmT(t)+\beta_\rmT \mathbf{s}^\rmH (t))}{\partial t} \right]\rmd t,\\
&=\int_{0}^{T_0}\mathbb{E}\left[(\alpha_\rmT^2+\beta_\rmT^2)\frac{\partial\mathbf{s}(t)}{\partial t}\frac{\partial\mathbf{s}^\rmT(t)}{\partial t}+2\Re\left\{\alpha_\rmT\beta_\rmT\frac{\partial\mathbf{s}(t)}{\partial t}\frac{\partial\mathbf{s}^\rmH(t)}{\partial t}\right\}\right]\rmd t,\\ 
&=N_\rms (8\pi^2)\alpha_\rmT\beta_\rmT W^2_\mathrm{eff}\mathbf{I}_{N_\mathrm{B}}=4\pi^2 N_\rms (1-m_\rmT^2\rme^{j2\psi_\rmT}) W^2_\mathrm{eff}\mathbf{I}_{N_\mathrm{B}}.
\end{align*}
Finally,
\begin{align*}
\int_{0}^{T_0}\mathbb{E}\Big[\frac{\partial\mathbf{s}_\rmT(t-\tau)}{\partial\tau}\mathbf{s}^\rmH_\rmT(t-\tau)\Big]\rmd t
=\int_{0}^{T_0}\mathbb{E}\Big[&\frac{\partial(\alpha_\rmT \mathbf{s}(t)+\beta_\rmT \mathbf{s}^* (t))}{\partial t}(\alpha_\rmT^* \mathbf{s}^\rmH(t)+\beta_\rmT^* \mathbf{s}^\rmT (t))\Big] \rmd t,\\
=\int_{0}^{T_0}\mathbb{E}\Big[&|\alpha_\rmT|^2\frac{\partial \mathbf{s}(t)}{\partial t}\mathbf{s}^\rmH(t)+|\beta_\rmT|^2\frac{\partial \mathbf{s}^*(t)}{\partial t}\mathbf{s}^\rmT(t)\\&+
\alpha_\rmT\beta_\rmT^*\frac{\partial \mathbf{s}(t)}{\partial t}\mathbf{s}^\rmT(t)+\alpha_\rmT^*\beta_\rmT\frac{\partial\mathbf{s}^*(t)}{\partial t}\mathbf{s}^\rmH(t)\Big] \rmd t=\mathbf{0}.
\end{align*}
Similarly, it can also be shown that $\int_{0}^{T_0}\frac{\partial\mathbf{s}_\rmT(t-\tau)}{\partial\tau}\mathbf{s}^\rmT _\rmT(t-\tau)\rmd t=\mathbf{0}$.

\bibliographystyle{IEEEtran}

\end{document}